\documentclass{mn2e}
\usepackage{epsfig}

\def\apj{ApJ}            
\def\mnras{MNRAS}        
\def\nat{Nature}         
\def\aa{A\&A}            
\newcommand{\phd}{\phantom{0}}
\newcommand{\php}{\phantom{.}}
\newcommand{\phm}{\phantom{-}}
\newcommand{\phl}{\phantom{l}}
\newcommand{\ts}{\thinspace}
\newcommand{\phT}{\phantom{1}}
\newcommand{\Delimv}{\mbox{$\overline{\Delta \Phi}$}}
\newcommand{\Delim}{\mbox{$\Delta \Phi$}}
\newcommand{\Delimdot}{\mbox{$\dot{\Delta \Phi}$}}
\newcommand{\bA}{\bf \rm A}
\newcommand{\bB}{\bf \rm B}
\newcommand{\tAej}{\mbox{$\php_{\bA} t_{\rm ej}$}}
\newcommand{\tBej}{\mbox{$\php_{\bB} t_{\rm ej}$}}
\newcommand{\The}{\mbox{$\overline{\theta}$}}
\newcommand{\Theal}{\mbox{$\theta_{\alpha}$}}
\newcommand{\Thedel}{\mbox{$\theta_{\delta}$}}
\newcommand{\Tlao}{\mbox{$\tilde{a_1}$}}
\newcommand{\Tlaf}{\mbox{$\tilde{a}_4$}}
\newcommand{\Tlbo}{\mbox{$\tilde{b}_1$}}
\newcommand{\Tlbf}{\mbox{$\tilde{b}_4$}}
\newcommand{\Tlp}{\mbox{$\tilde{p}$}}
\title{Helical Jet in the Gravitationally Lensed Blazar PKS1830-211}
\author[S. Nair et al.]
{S. Nair$^1$, 
C. Jin$^2$ \& M.~A.~Garrett$^3$
\\ \phd$^1$ Astronomy \& Astrophysics Group, Raman Research
Institute, C. V. Raman Avenue, Bangalore 560080, India \\
\phd$^2$National Astronomical Observatories, Chinese Academy of Sciences, 
A20 Datun Road, 
Chaoyang District, Beijing 100012, China\\
\phd$^3$Joint Institute for VLBI in Europe, NFRA, 7990 AA Dwingeloo,
The Netherlands  \\}

\date{\it Accepted: June 27, 2005 \quad Received: May 6, 2005 \quad In original form:
March 11, 2005}

\begin{document}
\label{first page}

\maketitle
\vskip 3mm
\begin{abstract}

Recent radio VLBI observations of the complex gravitationally lensed
system PKS1830-211 have thrown up some questions with regard to the
processes occuring at the heart of the blazar source at a redshift of
2.51, which is viewed almost straight down the jet axis.  This work
links, by a model of a helical jet tracked by ballistically ejected
plasmons from a precessing nozzle, observations on the scale of tens
of microarcseconds to those on the scale of milliarcseconds.  An
observed jet precession period of 1.08 years is inferred from the
model, translating to an intrinsic period of 30.8 years for a source
at redshift $z_s=2.51$ and an assumed jet bulk velocity $\beta$ of
0.99$c$. This fits well with the picture of the active galactic
nucleus hosting a binary black hole system at its centre, with the jet
emitted by one member of the system, and the precession as being due
to its orbital motion around its companion.

\end{abstract}


\section{ Introduction}
\label{intro}

The much-studied gravitationally lensed system, PKS1830-211, was
identified as such by Rao \& Subrahmanyam (1988), and is one of the
strongest radio sources in the sky at centimetric wavelengths. On
arcsecond scale, it consists of two prominent radio core-knot
structures, lensed images of each other, each core-knot axis being
roughly perpendicular to the line of separation of the two cores (the
two cores being separated by about $0^{"}.98$ and aligned roughly
northeast (NE) and southwest (SW) in the plane of the sky). The two
core-knot groups (hereafter NE and SW groups) are point-inversion
symmetric with respect to the centre of the system, and show somewhat
weaker diffuse emission (Rao \& Subrahmanyam 1988, Subrahmanyam et
al. 1990); in the radio L-band, the diffuse emission is extended
enough that its images actually close round, linking the image groups
and making a pseudo-Einstein Ring system (Jauncey et al. 1991), with
nearly half of the total flux density of about $12$ Jy coming from the
cores.The cores themselves have flat radio spectra (Rao \&
Subrahmanyan 1988). On the scale of milliarcseconds in the radio at 43
GHz, a weaker, inner knot accompanies each of the bright cores, on the
same side of each core as the arcsecond-scale knot (which is not
detected at this resolution), and separated from it by $\sim 1$ mas
(Garrett et al 1997, Jones et al., Guirado et al. 1999, Jin et
al. 2003).  

Jin et al. (2003) infer that there are time-dependent
positional variations in the radio centroids of emission in the bright
cores at 43 GHz, resulting in changes in the separation of the two
core components on the scale of $10$'s to $100$'s ${\mu}$as
(Figure~\ref{linear}) between successive observational epochs.
Significant radio variability has been reported (Lovell et al. 1996,
van Ommen et al.  1995, Hagiwara et al. 1996 and Lovell et al. 1998);
in fact, Lovell et al.(1998) report a time delay between the arrival
time at the observer of variations in one core image relative to
(correlated) variations in the other, of $26^{+4}_{-5}$ days. However,
Romero et al.(1997) report this system to be non-variable, based on
data taken during 1995-1996 in the radio L-band. Lovell et al's (1998)
analysis was based on $8.6\;$ GHz data between 1996 and 1997, during
which period dramatic variations occured.

The source has been identified in EGRET observations (Mattox et al., 1997).

In the X-ray, ROSAT obseravtions have revealed absorption against the
compact images, in excess of what may be reasonably attributed to our
Galaxy, implicating a foreground (lensing) galaxy (Mathur \& Nair
1997). ASCA observatons (Oshima et al. 2001) suggest significant X-ray
variability in the system, so that the ratio of the NE image's
emission to that of the SW image was a factor of $7$ (as compared with
a typical factor of $1$ to $1.8$ in the radio). Given that the source
properties resemble those of a blazar, this is very likely a variation
intrinsic to the blazar rather than a lens-induced one, though
observations of correlated spectral hardness changes would be
necessary to confirm that such a phenomenon indeed occurs in this
system.

The system that acts as the gravitational lens, now known to be fairly
complex, is believed to be in the main a gas-rich massive spiral
galaxy, with a measured redshift of $z_{1}=0.89$, through molecular
absorption line studies (Wiklind \& Combes 1996, Carilli et al. 1998,
Chengalur et al.  1999).  In the optical, observations are hindered by
the location of the system (in Galactic coordinates,
$l=12^o.2,\;b=-5^o.7$; here optical extinction is experienced of order
$2.7$ magnitudes, $cf.$ Subrahmanyan et al 1990); nonetheless Frye et
al. (1999)(I and K band observations), Courbin et al. (1998, 2002) (V,
I, J, H \& K bands), Lehar et al.(2000) (H and I bands) and Winn et
al. (2002)(I and V bands) have presented a wealth of optical and
near-IR data, which, suitably image processed, reveal a spiral galaxy
lens seen almost face on, with possibly a secondary lens nearby, and
perhaps some sign of a lower redshift spiral galaxy about $2^{"}.5$
southwards of the system. Lidman et al. (1999) determine through
infrared spectroscopy a redshift for the source $(z_s=2.51)$.  HI
observations (Lovell et al. 1996) have earlier revealed some
absorption at redshift $z_{2}=0.19$.  The presence in the optical and
IR bands of a foreground low redshift spiral galaxy is highly
suggestive of the existence of a secondary lens.

Gravitational lens modeling of the system that has attempted to
reproduce a range of image properties (Subrahmanyan et al. 1990,
Kochanek \& Narayan 1992, Nair et al. 1993), has shown the system to
be consistent with an isolated elliptical lens of mass of the order of
that of a spiral galaxy. However, in view of the apparently complex
optical/IR structure subsequently observed along the foreground and
upto the lens redshift, a detailed remodeling is in order.

The present work seeks to provide an explanation for the puzzling
observations of Garrett et al.(1998) and Jin et al.(2003), wherein it
is shown that structural and temporal variations occur in and around
the radio core images in this system in VLBA observations at 43
GHz. These observations probe the system at the highest resolution to
date, of considerable interest because of the combination of a high
redshift source $(z_s=2.5)$ and enhanced resolution due to
magnification by lensing.  For the purpose of the present study, a
full reworking of the lens model is actually inessential; instead,
certain lens properties applicable to this system will be either
invoked or derived as required.

\section{Some properties of the lens}
\label{lensmod}

Considering the optical and infrared observations of possible multiple
lensing bodies along the line of sight to the system (as described in
the previous section), only the most elementary assumptions regarding
the lens system will be made.  At the location of each image in the
plane of the sky, it is assumed that it is possible to define a $2
\times 2$ source-to-image coordinate transformation matrix that is
symmetric. This is valid provided one or the other of two
circumstances occurs:\ts (a)~the dominant lensing agents are roughly
at the same distance from the observer, or from the source, so that
they can be said to inhabit the same lens plane {\it (cf.} Blandford
\& Narayan $1986 )$, or (b)~save for the main deflectors in the same
lens plane, lenses at other redshifts provide only a shear
perturbation (however strong or weak) to the main lens's action. Then
the lensing equations reduce to one identical with single-plane
lensing, and source-to-image transformation matrices are symmetric as
a result (Kovner 1987).  The latter assumption is adopted in view of
the possible existence of a lens at $z=0.19$, which is seen $2.^"5$
south, in the plane of the sky, of the lensed images in the system. In
fact, since it has proven fairly successful to model this system with
a single elliptical lens (Subrahmanyan et al. 1990, Kochanek \&
Narayan 1992, Nair et al. 1993), it is possible that the main lens, a
spiral galaxy at $z_{1}=0.89$ which is viewed almost face-on, is aided
by shear roughly along the north-south direction by the foreground
lens, mimicking the behaviour of a single elliptical lens with such an
orientation (note that this is close to the direction of the major
axis of the elliptical lens model in Nair et al. 1993). That the
effect of the foreground $(z_{2}=0.19)$ lens can be treated as a
perturbation on the main lens' action derives some support from the
fact that lens models place the effective lens centre between the NE
and SW groups of images, and somewhat closer to the negative-parity SW
group of images (the parity of which is demonstrated by the fact that
temporal variations in its core's radio flux density lag behind those
in the NE core; $cf.$ Lovell et al. 1998). This would be as expected
for a single lensing galaxy, suggesting that the actual lens is not
very far from one.

\section{ A lensed helical jet}
\label{heljet}

\subsection{Inferred source evolution from observations 
 on the scale of tens of microarcseconds}
\label{microjet}

\begin{figure}
\begin{center}
\epsfig{file=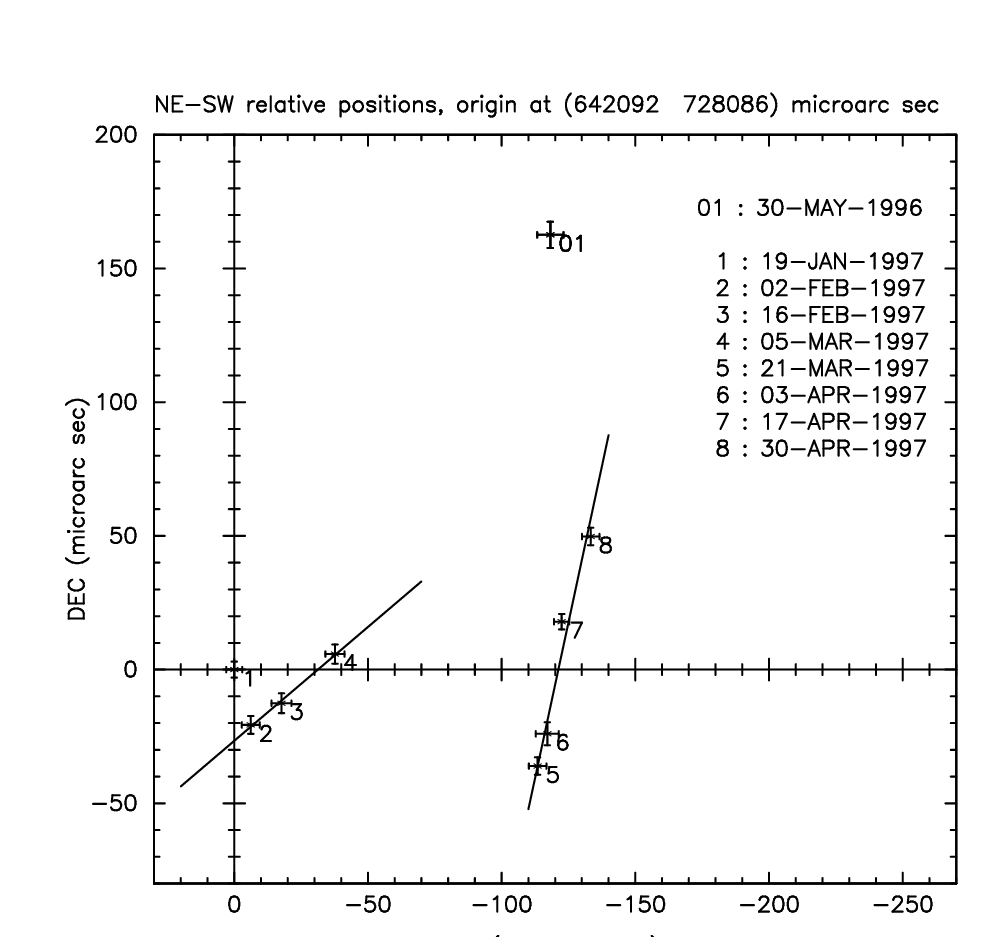,width=7.8 cm,angle=-90}
\end{center}
\caption{\label{linear}}{Changes in the centroids of radio emission in
the NE and SW cores; their separations as a function of epoch relative
to Epoch 1 in the figure (from Fig.3 of Jin et al. 2003). Superposed
are straight line fits to the two linear runs apparent in the data,
which yield slopes of (for epochs 2-3-4) $0.85 \pm 0.07$ and (for
epochs 5-6-7-8), $4.66 \pm 0.48$.}
\end{figure}

Figure~\ref{linear} (from Fig.3 of Jin et al. 2003), plots, for 8
successive epochs of VLBA observations at 43 GHz $(0.^{"}3$ mas FWHM
beam), changes in declination versus changes in right ascension (RA)
of the position vector from the centroid of emission of the NE image
core to that of the SW image core, all relative to the first epoch
($t_{\rm ref}$) of observation. Formally, the error in estimating
these changes is of the order of $3\;\mu$as as both the core images
appear in the same field of view in each observation (Jin et
al. 2003). The plot is displayed here with straight-line fits to two
remarkably linear tracks that can be identified in the plot. In this
section, an explanation for this phenomenon will be sought.

Anticipating a physical model to be discussed later in this section,
it is assumed that the source, a blazar, emits plasmons, one by one,
from its central engine, and that the observed (but essentially
unresolved) core radio emission is dominated for a while by that of
the most recently emitted plasmon, which propagates away from the
central engine with constant velocity before fading from view (its
spectral evolution, synchrotron self-absorbed, being as described in,
for example, van der Laan 1966). Any changes in the location of the
radio emission of the central engine itself are assumed to be
undetectable as a function of time $(\leq 3\; {\mu}$as).  (The actual
position of the central engine then drops out of the following
computations, because we consider changes in position relative to a
reference epoch).

The radio core of the source and its associated features are lensed
into two images, denoted by the subscripts $\bA$ (the NE group) and
$\bB$ (the SW group) in what follows. Changes occuring in the unlensed
source shall be referred to as being in the {\it source plane}, and
the corresponding effects in the images shall be referred to as
occuring in the {\it image plane}.  For the unlensed source, let the
angular distance of an ejected plasmon from the central engine at any
given epoch of observation, $t_k$, be denoted by $\Delimv[t_k]$, its
components along the directions of increasing RA and declination being
represented by $\Delim_{\alpha}$ and $\Delim_{\delta}$ respectively.

Now, local to a point in a gravitationally lensed image, positional
changes are mapped through a linear transformation of the underlying
changes in the source. That is, for small changes $\Delimv$ in the
source, the corresponding changes in images A and B are given by:

\begin{equation} \big(\Delim_{\alpha,\bA},\;\; \Delim_{\delta,\bA}\big) =
 \left( { \begin{array}{cc}
           a_1 & p\phl \\
           p\phl   & a_4 \\
          \end{array} }  \right)  \left( \!\! {\begin{array}{c}
                                                 \Delim_{\alpha} \\
                                                 \Delim_{\delta} \\
                                                 \end{array}} \! \! \right) 
\label{eq:tmtxA}
\end{equation} 

\begin{equation} \big(\Delim_{\alpha,\bB},\;\; \Delim_{\delta,\bB}\big) =
\left( { \begin{array}{cc}
           b_1 & q\phl \\
           q\phl   & b_4 \\
          \end{array} } \right) \left( \! \! {\begin{array}{c}
                                                 \Delim_{\alpha} \\
                                                 \Delim_{\delta} \\
                                                 \end{array} } \! \! \right) 
\label{eq:tmtxB}
\end{equation}

Two comments are in order at this point:$\ts$ (a)~It has been assumed
that the source-to-image transformation matrices are symmetric, under
the circumstances discussed in Section~\ref{lensmod} and (b)~, when
considering observed changes in the images, it is necessary to account
for the fact that events in the SW image are behind those in the NE
image, by an amount equal to the time delay between the images, $\tau
= {26}^{+5}_{-4}$ days (Lovell et al. 1998). Thus the observed changes
in the two images map back to two different epochs in the source
itself (epochs referring to events in the source will hereinafter be
called {\it source epochs}, as distinct from {\it observer epochs}).

Formally, it is possible to define a relative or image-to-image
transformation matrix (in this case, the SW image-to-NE image, or
`SW2NE', matrix will be defined):

\begin{equation} \big(\Delim_{\alpha,\bA},\;\;\Delim_{\delta,\bA}\big) =
 \left( { \begin{array}{cc}
                T_1 &  T_2 \\
                   T_3 & T_4 \\
                  \end{array} } \right) \left( \!\! {\begin{array}{c}                                              \Delim_{\alpha,\bB} \\
                                                   \Delim_{\delta,\bB} \\
                                                   \end{array} } \! \! \right) ,
\label{eq:sw2ne}
\end{equation}

\noindent where $\eta = (b_1b_4 - q^2)$, $T_1=(a_1b_4 -pq)/{\eta}$,
$T_2=(b_1p-a_1q)/{\eta} $, $T_3=(pb_4-qa_4)/{\eta}$, and $T_4=(a_4b_1
-pq)/{\eta}$. When evaluating this matrix from the observations, which
here include evolving components, care must be taken that appropriate
corrections are introduced for the differing source epochs (or
alternatively, light arrival time delays) corresponding to each of the
images.

\begin{figure}
\begin{center}
\epsfig{file=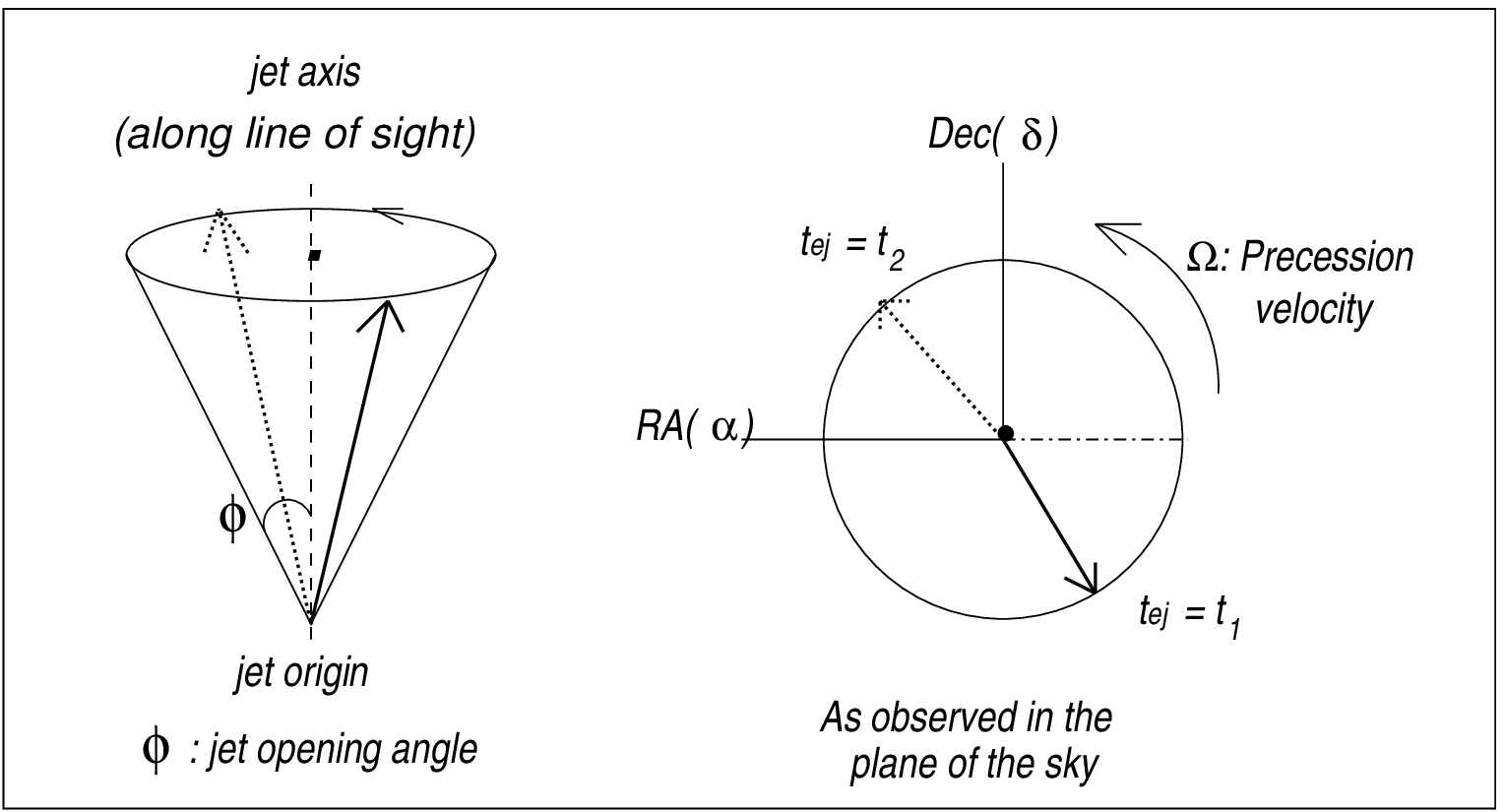, width=8.4 cm, angle=0}
\end{center}
\caption{\label{js}}{Underlying model for the precessing jet. The jet
emission vector makes an angle $\phi$ (jet half-opening angle) with
its axis, which is also the line of sight by assumption {\it (left)}.
From time to time, plasmons are ejected, and each propagates with
constant velocity along the surface of the cone defined by the motion
of the emission vector. {\it (Right):} The view along the line of
sight to the blazar, prior to lensing.}
\end{figure}

Adopting the Hjellming \& Johnston (1981) model for SS433 (or the
Gower et al. 1982 model, as applied to AGNs) for a precessing helical
jet, we assume that a plasmon in the jet is ejected from the central
engine of the source with constant velocity $\overline{v}$ along the
surface of a cone of half-opening angle $\phi$, the symmetry axis of
which is the axis of the jet ($cf.$ Fig.\ref{js}). New plasmons are
ejected ballistically from time to time.  The plasmon ejection
velocity vector precesses with angular velocity $\Omega$ about the
cone's surface, making a constant angle $\phi$ with the jet axis, so
that each subsequent plasmon is ejected at a different phase of the
jet's precession.  Once ejected with a given velocity $\overline{v}$,
a plasmon continues to propagate away from the central engine with
that velocity (or it fades from view before it appears to decelerate
or accelerate). Each plasmon is assumed to consist of an isotropically
expanding cloud of synchrotron-emitting electrons, so that the
centroid of emission is not affected by its evolution. As mentioned
earlier, it is assumed that only one plasmon at a time dominates the
unresolved radio emission from the radio core.  It is also assumed
that the line of sight of the observer coincides with the jet
precession axis, for simplicity of calculations.

Let a plasmon be ejected from the central engine of the source.
The observable proper motion of the unlensed plasmon from the time of its
ejection, $t_{\rm ej}$, to the epoch, $k$, of observation, $t_k$,
would be given by its components in the plane of the sky:

\begin{equation} \Delim_{\alpha} = {v_{\alpha} (t_k - t_{\rm ej})\over{d\php 
(1 - v_x/c)}}
\label{eq:propalpha}
   \end{equation}

\begin{equation} \Delim_{\delta} = {v_{\delta} (t_k - t_{\rm ej})\over{d\php 
(1 - v_x/c)}}
\label{eq:propdelta}
   \end{equation}

\noindent Here $d$ is the angular diameter distance to the source, and
$c$ is the velocity of light. The source time interval $(t_k -t_{\rm
ej})$ is corrected for the motion of the plasmon at relativistic
speeds along the line of sight during its propagation. (To convert a
source time interval itself to the observed time interval, $(\delta
t)_{\rm observer} = (\delta t)_{\rm source}/(1 -v_x/c)/(1+z_s)$, where
$z_s$ is the redshift of the source; here one corrects for
cosmological effects as well. This term is included in the angular
diameter distance in Eqns. \ref{eq:propalpha} \& \ref{eq:propdelta}
and therefore need not be specified explicitly).

At epoch $t_k$, we observe in image $\bA$:

$$\Delim_{\alpha, \bA}\php[t_k] = a_1\Delim_{\alpha}^{\bA} + p\phl\Delim_{\delta}^{\bA} $$
$$\Delim_{\delta, \bA}\php[t_k] = p\phl \Delim_{\alpha}^{\bA} + a_4 \Delim_{\delta}^{\bA}$$

\noindent where the superscript `$\tiny{\bf A}$' is used to indicate
that an appropriate correction for the light travel time along the
path through image $\bA$ must be applied. For image $\bB$:

$$\Delim_{\alpha, \bB}\php[t_k] = b_1\Delim_{\alpha}^{\bB} + q\phl\Delim_{\delta}^{\bB} $$
$$\Delim_{\delta, \bB}\php[t_k] = q\phl \Delim_{\alpha}^{\bB} + b_4 \Delim_{\delta}^{\bB}$$

\noindent with a similar use of superscript as before. The convention
that will be used in the following is that for events in image $\bA$,
$(t_k - t_{\rm ej})$ is written as $(t_k - \tAej)$, and for image
$\bB$, as $(t_k - \tBej) = (t_k -\tAej-\tau)$, explicitly writing the
light travel time difference $\tau$ between the paths through images
$\bA$ and $\bB$ into the expressions.

At this point contact is possible with the observations in
Figure~\ref{linear}.  What is observed in each image is a change
$\Delimv$ from a location $\The = (\Theal,\Thedel)$, the position of
the lensed central engine. It is assumed that $\The$ for each image
remains fixed through the different epochs of observation; since the
changes in Figure~\ref{linear} are plotted relative to a reference
epoch, the location $\The$ in each image will drop out of all
expressions thereafter, hence it will be ignored.

With this simplification, and making the appropriate substitutions,
the following quantities are constructed:

\begin{eqnarray} 
{\delta\Phi}_{\alpha}\php[t_k] &  = & \Delim_{\alpha, \bA}\php[t_k] - \Delim_{\alpha,
 \bB}\php[t_k] \nonumber \\
& = & \{(a_1 -b_1)v_{\alpha}^{\prime} + (p -q)v_{\delta}^{\prime}\}(t_k-\tAej) + \nonumber \\
\label{eq:dphialpha}
& &  (b_1 v_{\alpha}^{\prime} + q\phl v_{\delta}^{\prime})\tau  \\
\nonumber
{\delta\Phi}_{\delta}\php[t_k] &  = & \Delim_{\delta, \bA}\php[t_k] - \Delim_{\delta,
 \bB}\php[t_k]  \nonumber \\
& = & \{(p-q)v_{\alpha}^{\prime} + (a_4 -b_4)v_{\delta}^{\prime}\}\php(t_k-\tAej) + \nonumber \\
\label{eq:dphidelta}
& &  (q\phl v_{\alpha}^{\prime} + b_4 v_{\delta}^{\prime})\tau \\
\nonumber
\end{eqnarray}
 
\noindent In the above, the prime on quantities $v_{\alpha}$ and
$v_{\delta}$ is used to denote the corresponding velocity components
divided by $d\php (1-v_x/c). $ Also note the implicit assumption that
at the epoch of observation, $t_k$, the plasmon responsible for the
observations is present in both images (i.e., $t_k \ge \tAej + \tau).$
As long as only one plasmon at a time dominates the source emission on
this scale of observation, and it expands isotropically (so that the
centroid of radio emission is not affected by its evolution), it is
possible to neglect the effects of evolution of the plasmon itself
over the timescale of observation and deal simply with positional
changes of its centroid relative to the core. The presence of a second
evolving plasmon plasmon coexisting with the first would considerably
complicate this simple picture!

The changes plotted in Figure~\ref{linear} are relative to a reference
epoch $t_{\rm ref}$, so we construct new quantities, which are now the
abcissas and ordinates of the plot in Figure~\ref{linear} :
\begin{eqnarray}
\delta(\delta\Phi_{\alpha})\php [t_k] & = & {\delta\Phi}_{\alpha}\php[t_k] -
   {\delta\Phi}_{\alpha}\php[t_{\rm ref}] \nonumber \\
& = & (a_1 - b_1)\php v_{\alpha}^{\prime} +(p - q)\php v_{\delta}^{\prime} \\
\delta(\delta\Phi_{\delta})[t_k] & = & {\delta\Phi}_{\delta}\php[t_k] -
   {\delta\Phi}_{\delta}\php[t_{\rm ref}] \nonumber \\
& = & (p -q)\php v_{\alpha}^{\prime} +(a_4 -b_4)\php v_{\delta}^{\prime} \\
\end{eqnarray}

The slope $m$ of a particular linear track in Figure~\ref{linear} is
given by (dropping the primes on $v_{\alpha}$ and $v_{\delta}$ as
unnecessary here):
\begin{eqnarray}
 m & = & {\delta(\delta\Phi_{\alpha})\php[t_k] -\delta(\delta\Phi_{\alpha})\php[t_j]
}\over{\delta(\delta\Phi_{\delta})\php[t_k] -\delta(\delta\Phi_{\delta})\php
[t_j]} \nonumber \\
& = & {(a_1 -b_1)\php v_{\alpha} + (p\phT -q\phT )\php v_{\delta}}
\over{(p\phT - q\phT)\php v_{\alpha} + (a_4 -b_4)\php v_{\delta}} \\
\end{eqnarray}

From the above expressions it is now apparent that the change in slope
of the tracks between the run of epochs 2, 3 \& 4 (slope $m_1 =0.85
\pm 0.07$) and that of epochs 5, 6, 7 \& 8 (slope $m_2=4.66 \pm 0.48$)
is effected by a change in the quantities $v_{\alpha}$ and
$v_{\delta}$, projections of the ejection velocity vector
$\overline{v}$ of a plasmon, in the plane of the sky.  In the context
of our model, this corresponds to the dominance of a new plasmon
between epochs 4 \& 5, which has been ejected at a phase in the
precession of the jet which is distinct from that of the previously
ejected plasmon.  An obvious question at this point is, what is the
actual event that causes a discontinuity between epochs 1 \& 2, and
again between epochs 4 \& 5?  The appearance of a new plasmon in the
NE image precedes that in the SW image by 26 days, or nearly two
successive epochs in our 43 GHz observations, so if the discontinuity
were to be caused by its appearance in the NE image, there should be
another such discontinuity two epochs later, caused by its
manifestation in the SW image. The fact that both linear tracks are
more than two epochs long suggests that this is not the case. The
reason is somewhat subtle and will be discussed in
Section~\ref{justne} after deriving some quantities in connection with
the source-to-image transformation matrices, but for the moment we
identify the discontinuities in these linear tracks as corresponding
to the appearance of new plasmons in the {\it SW} image (so that, by
the time a discontinuity occurs, the {\it same} source plasmon is
imaged in both the NE and the SW regions).

As an aside, notice that for equal intervals between epochs, the changes
$\delta{\overline{\delta\Phi}}$ should be the same, along a single linear
track. That this is only approximately true for each of the two tracks in
Figure~\ref{linear} is an indication that emission near the central
core is variable at some level, pulling in or pushing out the
centroid of radio emission from time to time. This results in an 
apparent change 
in speed of the emergent plasmon, with no significant change in direction.

For later use, the above expression is recast in a slightly different form:
\begin{equation}
 {{v_\delta}\over{v_\alpha}}  = 
{{\Tlp(m\phT-\Tlao) - (m\phT-\Tlbo)} \over {\Tlp(1-m\Tlaf) - (1 -m\Tlbf)}}
\label{eq:tildad}
\end{equation}

\noindent where $\Tlp = p/q$, $\Tlao = a_1/p$, $\Tlaf = a_4/p$, $\Tlbo
= b_1/q$ and $\Tlbf = b_4/q$. For a given value of slope $m$, once the
lensing$-$related quantities $\Tlp,\;\Tlao,\;\Tlaf,\;\Tlbo\; \&
\;\Tlbf$ are determined, this expression gives the phase of precession
of the jet (modulo $n\pi$ radians, where $n$ is a natural number),
relative to a suitably chosen reference direction, at the time of
ejection of the relevant plasmon. Thus, the difference of the phases
between the events slopes $m_1$ and $m_2$, over the known difference
in epoch between the observations, yields the angular velocity or rate
of precession of the jet's source, $\Omega = \dot{\psi}$:

\begin{equation}
\Omega= {(\psi_2 -\psi_1)\over (t_2 -t_1)},
\end{equation}
\noindent where:
\begin{equation}
\psi_1=-Tan^{-1}\left( {v_{\delta} \over {v_{\alpha }}}[m_1] \right)
\nonumber
\end{equation}
\begin{equation}
\psi_2=-Tan^{-1}\left( {v_{\delta} \over {v_{\alpha}}} [m_2] \right)
\nonumber
\end{equation}
\noindent and $t_1$ and $t_2$ are just the dates of the first epoch of
observation of each event ($t_1$ is epoch 2, 2 Feb 1997, and $t_2$ is
epoch 5, 21 Mar 1997).  The uncertainty in the commencement of each
event is regarded as being uniformly distributed, between epochs 1 and
2 for $t_1$, and between epochs 4 and 5 for $t_2$ (see
Fig.~\ref{linear}).

\subsection{\label{milli} The milliarcsecond$-$scale behaviour}

\begin{figure}
\begin{center}
\epsfig{file=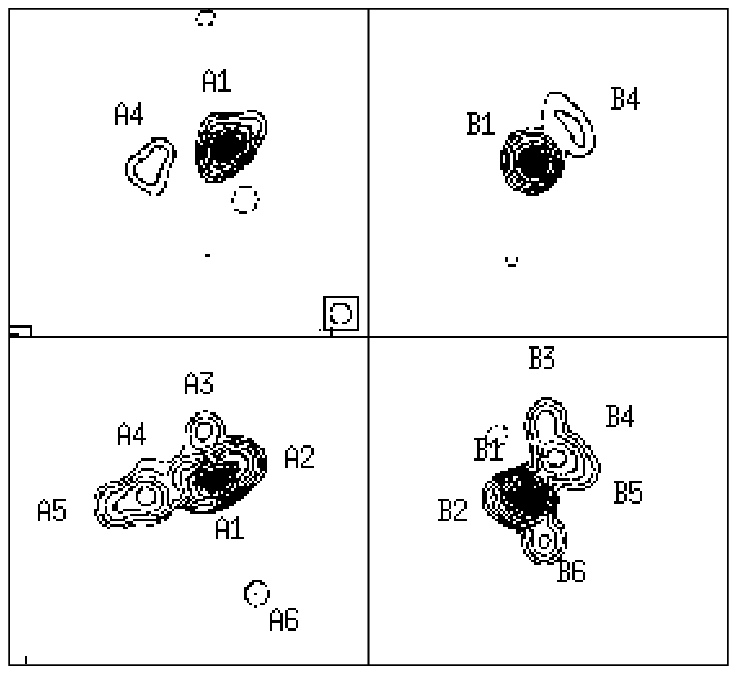,width=10.4 cm,angle=0}
\end{center}
\caption{\label{masplot}}{$(Top):$ The observations of 30 May 1997 of the NE $(left)$
and SW $(right)$ VLBA images at 43 GHz and 0.33 mas circular restoring beam, 
showing a simple core-knot
structure in each image. $(Bottom):$ The same as above, but for the next epoch 
of observations,  14 July 1996, exhibiting rich structure in both images.(Both
maps from Garrett et al., 1998.)}
\end{figure}

At this point, it is necessary to review some puzzling VLBA
observations of PKS1830-211 at 43 GHz which preceded the eight-epoch
track of observations of Jin et al. (2003), namely, those of Garrett
et al.(1998). These observations revealed dramatic changes in the
milliarcsecond scale structure of the two image groups between epochs
of observation 30 May 1996 and 14 July 1996. The former epoch of
observation showed core-knot structures in both the NE and SW images,
as for each of the epochs of observation in Jin et al.(2003), but the
latter epoch, quite suddenly and surprisingly, exhibited rich
structure in each of the image groups, with as many as six features in
the NE image group being potentially identifiable with corresponding
features in the SW image group (including the basic core-knot
structure, $cf.$ Fig.~\ref{masplot}).  By the time of the beginning of
Jin et al.'s (2003) observations (19 January 1997), the system was
back to its quiescent phase, with only a simple core-knot structure
being imaged. At some point, on a timescale of several weeks to half a
year, a dramatic outburst of plasmons occured, which then evolved and
faded from view, and did not recur over the three months that
followed, during which Jin et al.'s (2003) monitoring observations of
the system were in progress.

It is expected that at least some of the plasmons in the NE image will
have corresponding entities in the SW image. Obviously related are
features A1 and B1, the core images, and A4 and B4, which are seen in
both epochs of observation in Figure~\ref{masplot} and appear to be
static `knots'. Of the short-lived features observed in the maps of 14
July 1996, A3 and B3 are obviously not related to each other, from
considerations of image parity.  However, A2 and B2, A5 and B5, and A6
and B6 are easily seen to be corresponding images, once the effects of
the time delay between the NE and SW images (recalling that phenomena
in the NE group of images occur about 26 days earlier than they show
up in the SW image group), and the distorting effects of lensing in
the two image groups are accounted for.  It is conjectured at this
point that feature A3 has not as yet manifest a counterpart in the SW
group of images (it is younger than 26 days), and feature B3 is so
aged that its counterpart in the NE group has already evolved below
the map detection limits.

\subsection{Deriving the relative (SW2NE) image transformation matrix}
\label{transmatrix}
Assuming that the source-to-image linear transformation matrices
remain locally valid across the entire NE or SW image group (on scales
of order 1 mas), and that each plasmon in the source is moving
ballistically away from the core, on a straight line trajectory, it
proves feasible to derive a relative (image-to-image) $2 \times 2$
transformation matrix from the observations of 14 July 1996 shown in
Figure~\ref{masplot}.

Recasting the relative image transformation matrix in the notation of 
Equation~\ref{eq:tildad},
\begin{eqnarray}
 \big(\Delim_{\alpha,\bA},\;\; \Delim_{\delta,\bA}\big) &= &
            \left( {\begin{array}{cc} 
                   T_1   & T_2  \\ 
                   T_3   & T_4  \\
                   \end{array}} \right) \left( { \begin{array}{c}
                                                \Delim_{\alpha,\bB} \\
                                                \Delim_{\delta,\bB} \\
                                                \end{array}} \! \! \right),
\label{eq:tildadsw2ne}
\end{eqnarray}
\noindent with $T_1=\aleph(\Tlao\Tlbf-1)$, $T_2=\aleph(\Tlbo-\Tlao)$, 
$T_3=\aleph(\Tlbf-\Tlaf)$ and
$T_4=\aleph(\Tlaf\Tlbo-1)$, and $\aleph={\Tlp/{(\Tlbo\Tlbf-1)}}$.

For a given feature $i$, its separation from the core position will be
denoted by the vector $(\Delim_{\alpha,\bA_i},\Delim_{\delta,\bA_i})$.
From the static features, the cores A1 and B1, and knots A4 and B4,
one derives the core-knot separation vector in each image group,
yielding two constraints (positional coordinates) on the relative
transformation matrix (from Eqn.~\ref{eq:tildadsw2ne}):

\begin{equation}
T_1\Delim_{\alpha,\bB_4} + T_2\Delim_{\delta,\bB_4} -\Delim_{\alpha,\bA_4}=0
\label{eq:cons1}
\end{equation}
\begin{equation}
T_3\Delim_{\alpha,\bB_4} + T_4\Delim_{\delta,\bB_4} -\Delim_{\delta,\bA_4}=0
\label{eq:cons2}
\end{equation}

The pairs of relationships between A2 and B2, A5 and B5, and A6 and B6
each yield a single constraint, from the shared {\it direction} of
motion of the plasmon in the source (these are evolving features and
their actual locations in the NE and SW images refer back to different
source times due to the relative time delay). That is, given the
location of, say, plasmon B2 in the the SW group relative to its core
image B1, the relative transformation matrix must be such that it
reproduces the correct direction for plasmon A2 relative to its core
A1 in the NE group of images.

\noindent For features A2 and B2, A5 and B5, and A6 and B6, the single
constraint equation from each pair reads (with $i=2,5,6)$:

\begin{equation}
{\Delim_{\delta,\bA_i}\over{\Delim_{\alpha,\bA_i}}}= {T_3\Delim_{\alpha,\bB_i}+
T_4\Delim_{\delta,\bB_i} \over {T_1\Delim_{\alpha,\bB_i}+T_2\Delim_{\delta,\bB_i}}} 
\nonumber
\end{equation}

\noindent which may be rewritten as:

\begin{eqnarray}
T_1\Delim_{\alpha,\bB_i}\Delim_{\delta,\bA_i} +  T_2\Delim_{\delta,\bA_i}\Delim_{\delta,\bB_i} -& &
\nonumber \\
T_3\Delim_{\alpha,\bA_i}\Delim_{\alpha,\bB_i} - T_4\Delim_{\alpha,\bA_i}\Delim_{\delta,\bB_i}
& =0&
\label{eq:cons3}
\end{eqnarray}

Thus the observations yield, through Equations~\ref{eq:cons1},
\ref{eq:cons2} and \ref{eq:cons3}, 5 constraint equations for the 4
unknown relative transformation matrix elements, $T_j,\;\; j=1$ to
$4$. Note that the unknown matrix elements and the observations
determining them are implicitly related, non-linear in the
observational quantities, and of the form: ${\bf F}({\bf X}_a,{\bf
L}_a)={\bf 0}$, where ${\bf X}_a$ is the vector of 4 unknowns (matrix
elements) and ${\bf L}_a$ is the vector of 16 observational
quantities. ${\bf F}$ itself consists of 5 nonlinear functions. The
unknowns in vector {\bf X} are solved for via a {\it mixed adjustment
model} (see, for example, A. Leick, 1995), the solution of which is
described in the Appendix.

The solution for the elements of the SW2NE transformation matrix are:

\begin{eqnarray}
  {\bf T}_{SW2NE}= &
            \left( {\begin{array}{cc}
                   -1.279   & 0.871  \\
             \phantom{+}0.973   & 0.245  \\
                   \end{array}} \right) 
\label{eq:sw2neval}
\end{eqnarray}

\noindent Its determinant is: $-1.161.$ This is the NE/SW image flux
density ratio, $k = T_1T_4 -T_2T_3$. The covariance (error) matrix
is given by:

\begin{eqnarray}
 {\bf C_T} = &
        \left( {\begin{array}{cccc}
          0.071 & 0.031 & 0.106 & 0.053 \\
          0.031 & 0.017 & 0.054 & 0.026 \\
          0.106 & 0.054 & 0.342 & 0.165 \\
          0.053 & 0.026 & 0.165 & 0.081 \\
                \end{array}} \right)
\label{eq:covmat}
\end{eqnarray}

\noindent From the expressions immediately following
Equation~\ref{eq:tildadsw2ne}, it is possible to express the
quantities $\Tlao$, $\Tlaf$, $\Tlbo$ and $\Tlbf$ in terms of $T_1$,
$T_2$, $T_3$ and $T_4$ and $\Tlp$.

\begin{equation}
\Tlao = (T_1 -k/\Tlp)/T_3 
\label{eq:atil1}
\end{equation}
\begin{equation}
\Tlaf = (T_4 -k/\Tlp)/T_2 
\label{eq:atil4}
\end{equation}
\begin{equation}
\Tlbo = (\Tlp -T_4)/T_3 
\label{eq:btil1}
\end{equation}
\begin{equation}
\Tlbf = (\Tlp -T_1)/T_2 
\label{eq:btil4}
\end{equation}

\noindent The evaluation of these quantities depends on the value of
$\Tlp$, for which straightforward algebraic methods of determination
exist, but which were found to be sensitive to numerical errors of
estimate. Hence a different approach has been taken, which will be
discussed in Section~\ref{evaltwo}.

\subsection{The directions of ejection for the milliarcsecond-scale features}
\label{dirmilli}

It is possible, under the explicit assumption that the source velocity
for each evolving plasmon remains constant from the time of generation
at least upto the epoch represented by the observations of 14 July
1996 in Figure~\ref{masplot}, to relate the distance travelled by a
given plasmon as seen in the NE image group of that observation, with
the distance that is travelled by its corresponding feature,
transformed from the SW group of images via the relative image
transformation matrix ${\bf T}_{SW2NE}$ and corrected for the lag in
time of 26 days for the SW group.

That is, the squared distances are (for $i=2,5,6$):
\begin{equation}
{\rm D}_{NE}^2 = {\Delim^2_{\alpha,\bA_i}} +{\Delim^2_{\delta,\bA_i}},
\end{equation}
in the NE group of images, and transformed from the SW group to the NE
group, are:
\begin{eqnarray}
{\rm D}_{SW2NE}^2 =& {(T_1\Delim_{\alpha,\bB_i}+T_2\Delim_{\delta,\bB_i})}^2 + & \\ \nonumber
& {(T_3\Delim_{\alpha,\bB_i} + T_4\Delim_{\delta,\bB_i})}^2 &
\end{eqnarray}
\noindent In the NE group, the velocity of motion is ${\rm
D}_{NE}/(t_{k}-t_{ej})$, and, transformed from the SW group, the
corresponding velocity of motion is ${\rm D}_{SW2NE}/ (t_k-t_{ej}
-\tau)$, where $t_k$ is the epoch of observation and $\tau$ is the
time delay between the image groups (assumed to be constant over the
span of about a milliarcsecond from the core). If ${\cal Q}$
represents the ratio ${\rm D}_{NE}/{\rm D}_{SW2NE}$, then:

\begin{equation}
t_k-t_{ej} = {\tau {\cal Q} \over {{\cal Q} -1}},
\label{eq:ttrav}
\end{equation}
\noindent with ${\cal Q} > 1.$ 

In order to estimate how well the mixed adjustment model-derived
relative transformation matrix actually reproduces the position angles
of the features in the NE group of images when it is applied to data
from the SW group, the quantities
$\Delim_{\delta,\bA_i}/\Delim_{\alpha,\bA_i}$ (for the original
position angle of feature $i$ in the NE group), and
$(T_3\Delim_{\alpha,\bB_i} + T_4\Delim_{\delta,\bB_i})/
(T_1\Delim_{\alpha,\bB_i}+T_2\Delim_{\delta,\bB_i})$ (for the position
angle of the transformed data relating to feature $i$ from the SW
image) are calculated. For feature A2 (B2), the original position
angle is $-48^{\rm o}$ versus a transformed value of $-30^{\rm o}$ but
${\cal Q}$ is $<1$ (pointing to some inconsistency for this feature);
for feature A5 (B5), the original position angle is $104^{\rm o}$ as
compared with a transformed value of $113^{\rm o}$, and ${\cal Q}$ is
1.272; from Equation~\ref{eq:ttrav}, a value of 125.2 days is obtained
as the `age' of the feature at the time of observation in the NE image
on 14 July 1996. For feature A6 (B6), the original position angle is
$159^{\rm o}$, and the transformed position angle is $158^{\rm o}$;
${\cal Q}$ is 6.255 and the `age' of the feature as seen in the NE
image is 30.9 days.(Owing to the existence of significant
cross-correlations between quantities derived through the mixed
adjustment model -- $cf.$ Equation~\ref{eq:covmat} -- formal errors on
quantities derived will be given only for the final result). A
cross-check on the mapping of the static knot reveals a perfect match
for position. The position angle is $99^{\rm o}$ in the original;
transformed, it is $99^{\rm o}$, and as to the distance from the core,
it is 0.99 mas in the original and 0.99 mas in the transformed
version. Thus the transformation matrix is heavily influenced by
feature A6(B6) and the the location of the static knot feature.

Of course, in order to derive the rate of precession of the jet from
these numbers, it is necessary to consider the mappings in the source
plane of these features, rather than those in the NE image. For this,
it is necessary to calculate the velocities of the various evolving
features in the source plane. The elements of the source-to-image
transformation matrix in Equation~\ref{eq:tmtxA} can be rewritten,
dividing each element by $p$. Then its inverse reads:
\begin{equation} {\bf T}^{-1}_{S2NE} =
 {p^{-2}(\Tlao\Tlaf-1)^{-2}} \left( { \begin{array}{cc}
           \Tlaf & -1 \\
            -1   & \Tlao \\
          \end{array} }  \right)  
\label{eq:invtmtxAscal}
\end{equation}
\noindent Similarly, for Equation~\ref{eq:tmtxB}, the matrix can be
recast and inverted to yield:
\begin{equation} {\bf T}^{-1}_{S2SW} =
 {q^{-2}(\Tlbo\Tlbf-1)^{-2}} \left( { \begin{array}{cc}
           \Tlbf & -1 \\
            -1   & \Tlbo \\
          \end{array} }  \right)  
\label{eq:invtmtxBscal}
\end{equation}

\noindent Taking the time-derivative (denoted by a superscribed dot)
of the angular separations involved, so that one is now dealing with
velocities,
\begin{eqnarray}
{\Delimdot}_{\alpha} = &{p^{-1}(\Tlao\Tlaf-1)^{-1}}\left(\Tlaf{\Delimdot}_{\alpha,A} -
{\Delimdot_{\delta,A}} \right)& \\
=&{q^{-1}(\Tlbo\Tlbf-1)^{-1}}\left(\Tlbf{\Delimdot}_{\alpha,B} -{\Delimdot}_{\delta,B} \right)&
\end{eqnarray}
\begin{eqnarray}
{\Delimdot}_{\delta} = &{p^{-1}(\Tlao\Tlaf-1)^{-1}}\left(\Tlao{\Delimdot}_{\delta,A} -
{\Delimdot_{\alpha,A}} \right)& \\
=&{q^{-1}(\Tlbo\Tlbf-1)^{-1}}\left(\Tlbo{\Delimdot}_{\delta,B} -{\Delimdot}_{\alpha,B} \right)&
\end{eqnarray}
\noindent Now, from the inverse of the image-to-image transformation matrix,
\begin{equation}
\Delimdot_{\alpha,B}={k}^{-1}(T_4\Delimdot_{\alpha,A}-T_2\Delimdot_{\delta,A})
\end{equation}
\begin{equation}
\Delimdot_{\delta,B}={k}^{-1}(T_1\Delimdot_{\delta,A}-T_3\Delimdot_{\alpha,A})
\end{equation}
\noindent so that, substituting for $\Tlao$ and $\Tlaf$, or $\Tlbo$
and $\Tlbf$ from Equations~\ref{eq:atil1} to \ref{eq:btil4} and with a
little algebra, one obtains:
\begin{equation}
{\Delimdot_{\delta}\over{\Delimdot_{\alpha}}}={T_2 \over T_3}{(\Tlp\Delimdot_{\delta,\bB} -
\Delimdot_{\delta,\bA})\over {(\Tlp\Delimdot_{\alpha,\bB} - \Delimdot_{\alpha,\bA})}}
={v_\delta \over v_\alpha}
\label{eq:millidet}
\end{equation}
\noindent This expression yields (as in the microarcsecond-scale
 analysis discussed earlier) the phase (modulo $n\pi$) of ejection of
 the relevant plasmon, with a suitable choice of reference
 direction. The quantities $\Delimdot_{\alpha,\bB}$ and
 $\Delimdot_{\delta,\bB}$ are obtained from $\Delim_{\alpha,\bB}$ and
 $\Delim_{\delta,\bB}$ by dividing by $(t_k-t_{ej}-\tau)$, and
 $\Delimdot_{\alpha,\bA}$ and $\Delimdot_{\delta,\bA}$ are obtained by
 dividing $\Delim_{\alpha,\bA}$ and $\Delim_{\delta,\bA}$ by
 $(t_k-t_{ej})$.  To estimate the rate of precession of the jet from
 expression~\ref{eq:millidet}, the phases of ejection, $\psi_j$, are
 calculated for plasmons $j$ (with $j=5,6$) and the observed time that
 has lapsed between the ejection of plasmon 5 and plasmon 6 is
 calculated as $(t_k - t_{ej,6}) - (t_k-t_{ej,5})$, where $t_k$, the
 shared epoch of observation, drops out of the estimate. The rest of
 the calculation follows along the lines described for the
 microarcsecond scale analysis.

\subsection{Evaluating the period of precession of the jet from milliarcsecond 
and microarcsecond scale information}
\label{evaltwo}

Expressions~\ref{eq:tildad} and \ref{eq:millidet} are, respectively,
the phases of ejection of various plasmons on the microarcsecond and
milliarcsecond scales, as derived from observables. Note that both
expressions require the evaluation of $\Tlp$. In order to do this, it
will be assumed that the period of precession of the relativistic jet
remains the same when evaluated from data on the scale of tens of
microarcseconds, as from data on the scale of milliarseconds (for a
jet opening half-angle of, say, $0^{\rm o}.5$ and assuming a typical
linear magnification by the gravitational lens by a factor of 3, this
difference of scale corresponds to source-intrinsic times of order
$10^3$ years, too short to expect a significant evolution in the
precession velocity of the jet). Hence a range of values of $\Tlp$ is
considered, and its actual value is determined from the coincidence on
these two scales of calculated values of the rate of precession of the
jet, which exercise also determines the rate of precession itself.
   
\begin{figure}
\begin{center}
\epsfig{file=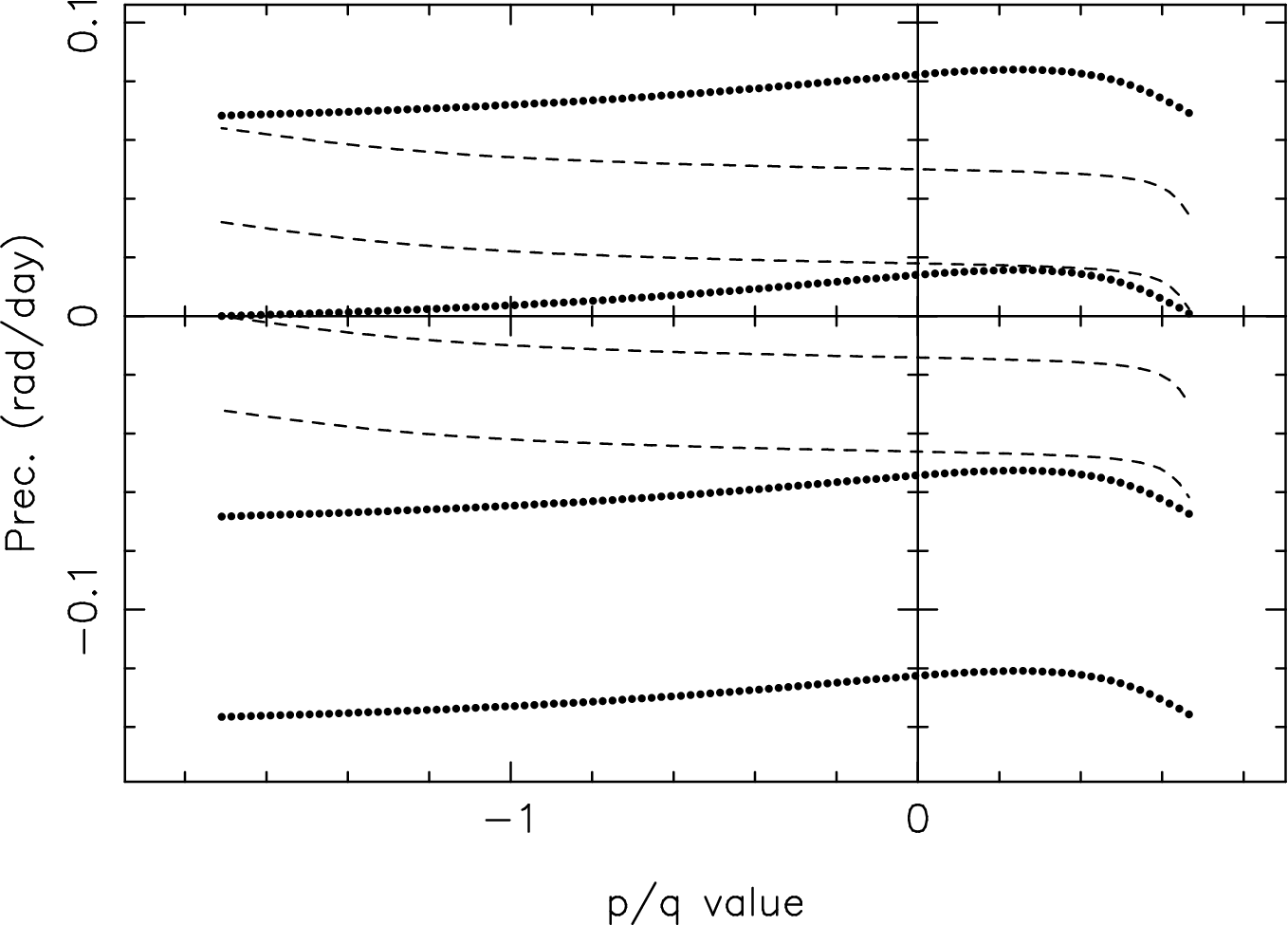,width=6.0 cm,angle=-90}
\end{center}
\caption{\label{branch4}}{Solution for the combination of the lensing
quantity $\Tlp$ and the rate of precession of the blazar jet (dots are
used for the microarcsecond scale solution, and dashed lines for the
milliarcsecond scale solution).  All four possible branches are
evaluated for each solution. The coincidence of the milliarcsecond
scale and the microarcsecond scale solution is the desired result
($p/q$=0.3, precession rate = 0.016 rad/day). For $p/q < -1.7$ and
$p/q > 0.7$, the NE image becomes negative parity, which is known to
be incorrect. Positive values of the precession rate are for
anticlockwise rotation in the source (as also for the NE image; parity
differences cause an opposite rotation in the SW image).}
\end{figure}

The rate of precession is evaluated as that which would have been
observed in the absence of the lens, rather than that intrinsic to the
source, since the component of the bulk velocity of the jet along the
line of sight cannot be determined (but is expected to be close to the
velocity of light).  It is calculated as described in
Sections~\ref{microjet} and \ref{dirmilli}, over a range of values
$\Tlp=p/q$. Recalling that the actual phases of ejection are only
determined modulo $n\pi$, there is a degeneracy in determining the
difference of phases $\psi_2-\psi_1$ for both the milliarcsecond and
the microarcsecond analyses. This means that each separation
$\psi_i-\psi_j$ actually corresponds to four possible branches. Each
of these branches is plotted against $\Tlp$ in
Figure~\ref{branch4}. At this point, it is apparent that a value of
$\Tlp$ of $\sim 0.3$, and a precession rate of $0.016$ radians/day
(corresponding to an observed period of 1.08 years) permit a
coincidence of the milliarcsecond and microarcsecond scale results, so
these values are adopted. To obtain the intrinsic source rotation
period, the observed period must be corrected for the cosmological
time dilation on account of the source being at a redshift of 2.5, and
for the change in time interval on account of the jet velocity in the
source. The first correction reduces the period by a factor of 3.5,
and the second increases it by a factor of $1/(1-\beta_{los})$, where
$\beta_{los} = v_x/c$, with c as the velocity of light.  For small jet
opening angles, $\beta_{los} \sim v_{jet}/c$.  Assuming a value of
$\beta=0.99$, the intrinsic period is calculated to be 30.8 years.

Deriving a period of 1.08 years from the observations raises a
question about the authenticity of the results obtained in this
paper. We have checked carefully for the possibility of effects
related to the annual motion of the Earth around the Sun.  The paper
of Jin et al. (2003), from which the microarcsecond scale observations
are drawn, included a correction for the differential annual
aberration between the NE and SW groups of images (separated by
$\sim\;1$ arcsec), which is typically of the order of tens of
microarcseconds. Relative annual parallax between the NE and SW core
images is of order $10^{-4}$ $\mu$as, too small to produce a spurious
effect in the observations.  (As an aside, another possible source of
lensing capable of causing a differential shift in the location of the
NE core image relative to the SW one is the Sun itself, which was
always more than 15 degrees away from PKS1830-211 in the observations
of Jin et al. (2003), thus producing a differential shift in the NE
core image relative to the SW core image of much less than a
microarcsecond). Another reason that we do not consider our present
result to be a consequence of the Earth's motion around the Sun is the
coincidence of this result as obtained from the milliarcsecond scale
data of Garrett et al. (1998) with that obtained from the
microarcsecond scale data of Jin et al. (2003), as shown in Figure
4. It is difficult to see how the annual motion of the Earth could
arrange for this, especially as the milliarcsecond scale data is just
one actual epoch of observation (which effectively includes two epochs
for the source only because of the existence of two groups of lensed
images mapping back to source events separated in time by the time
delay of 26 days).

The error in evaluating the precession period is estimated via a Monte
Carlo exercise that generates, for each of 60,000 experiments, a
random selection of all the observational errors, taking into account
the cross-correlations between elements of the SW2NE image-to-image
transformation matrix, and then computes the desired precession rate
for a value of $\Tlp=p/q$ of 0.3. For the microarcsecond scale
analysis, the precession rate (in radians per day with $1-\sigma$
errors) is $0.0155\;(-0.0166,0.0415)$, and for the milliarcsecond
scale analysis, the corresponding values are
$0.0164\;(-0.575,0.987)$. The large errors on the milliarcsecond scale
result are mainly due to the form of the expression used to evaluate
the times of ejection for the two plasmons (see Eqn.~\ref{eq:ttrav}).

The foregoing results are derived, as mentioned earlier, for an
assumed perfect alignment of the line of sight with the jet
axis. Typically, this is not the case, although it is expected that
the alignment of the two is to within a few degrees for blazar
sources. The effect of considering an inclination of the jet axis to
the line of sight can be estimated from expressions (1), (2) and (3)
for the $x$, $y$ and $z$ velocity components of a plasmon in the
precessing jet model of Gower et al. (1982), where axes $z$ and $y$
are in the plane of the sky and centred on the central engine, and
axis $x$ is along the line of sight. In general, an additional
rotation through an angle $\chi$ of Gower et al.'s $y$ and $z$ axes
about the $x$ axis is necessary to bring the model into alignment with
the right ascension and declination axes. Let the jet half-opening
angle be $\phi$, and let, as in earlier expressions, the angle through
which the jet has precessed at the time of observation, as measured
from a reference time $t_{ref}$, be $\psi$. Then, with $i$ as the
inclination angle of the jet axis to the line of sight, the inferred
value of $v_{\delta}/v_{\alpha}$ from, $e.g.$,
Equation~\ref{eq:millidet}, is related to the precession angle $\psi$
(for each plasmon considered) by:

\begin{equation}
{{Sin{\chi}+ ({v_{\delta}\over v_{\alpha}})Cos{\chi}}\over{Cos{\chi} - 
({v_{\delta}\over v_{\alpha}})Sin{\chi}}}
={{Sin{\phi}Sin{\psi}}\over{Sin(i)Cos{\phi}-Cos(i)Sin{\phi}Cos{\psi}}}
\label{eq:inclina}
\end{equation}

\noindent From the milliarcsecond scale data, for example, with a jet
inclination angle of $0^{\rm o}$, for feature 5, $\psi=200^{\rm o}$,
and $t_{ej}-t_{ref}= 125.2$ days, and for feature 6, the corresponding
values are $-69^{\rm o}$ and 30.9 days, yielding, as mentioned above,
a precession period intrinsic to the source of 30.8 years. For
non-zero values of $i$ and $\phi$, in order to obtain the precession
rate or period in the source rest frame, the timescales
$t_{ej}-t_{ref}$ must be in each case corrected by a factor of
$1/(1-v_x/c)$, where $v_x$ is the line of sight velocity:

\begin{equation}
{v_x \over c} = \beta \left( Sin{\phi}Sin(i)Cos{\psi} + Cos{\phi}Cos(i) \right)
\label{eq:vlos}
\end{equation}

\noindent With $i=8^{\rm o}$, $\phi=5^{\rm o}$, $\chi=60^{\rm o}$ and
$\beta=0.99$, the precession period in the source frame is 21 years. A
choice of $i=4.^{\rm o}5$ and $\phi=3^{\rm o}$ with $\chi$ and $\beta$
as before, yields a precession period of about 25 years.

\subsection{Single plasmon events on the microarcsecond scale}
\label{justne}

\begin{figure}
\begin{center}
\epsfig{file=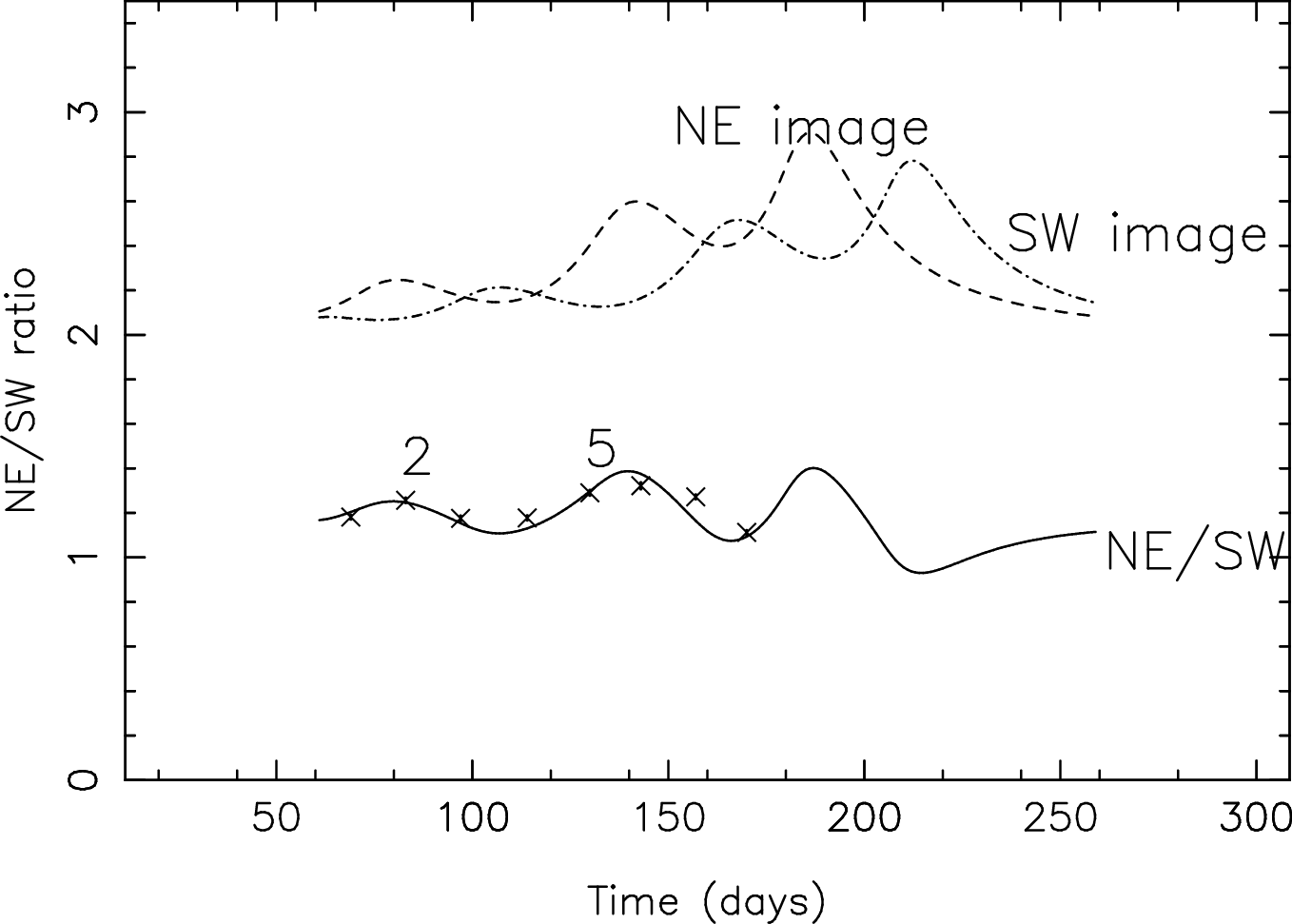,width=6.0 cm,angle=-90}
\end{center}
\caption{\label{plotratio}}{{\it(Solid curve):} Simulated variation of
the ratio of the NE image's flux density to that in the SW image at 43
GHz.{\it (Crosses):} 8 epochs of data, with errors within the extent
of the symbols used, from Jin et al. 2003).{\it (Dashed line):}
Simulated underlying flux density variation in the NE image core, and
{\it (dash-dot line):} corresponding time-delayed variation in the SW
image core. Epochs 2 and 5 of the data correspond to the commencement
of the observed microarcsecond-scale `events' discussed in
Section~\ref{microjet}.  }
\end{figure}

From Table 1 of Jin et al. (2003), the ratio of the NE to SW core
 image flux densities is seen to vary as a function of observational
 epoch ($cf.$ also the crosses plotted in Fig.~\ref{plotratio}).  The
 variation shows two apparent maxima, indicating source variability
 (although variations on the timescales of a couple of weeks are
 possible with a combination of an apparently superluminal source and
 microlensing by stars in the lens galaxy --- $cf.$ Gopal-Krishna \&
 Subramanian 1991 --- it will be post-justified that this is not the
 case in the present circumstance). It is tempting to model these
 variations in terms of the emission of synchrotron self-absorbed
 plasmons. Following Expression 11 of van der Laan (1966), a series of
 four succesively emitted plasmons is considered, expanding with
 constant velocity, with an initial spectral maximum at 43 GHz, and
 electron power law index of 3/2. Since the model is actually
 underconstrained by the observations, it is only attempted to seek a
 consistency between the two. For each image, the emission from the
 series of four plasmons is computed. In the case of the SW image,
 this curve is scaled by the NE/SW image flux density ratio $\mid k
 \mid = 1.161$, derived from Section~\ref{transmatrix}, and delayed in
 time relative to the NE curve by the time delay of 26 days.  The
 middle two plasmons are constrained to have an {\it observed}
 separation, in terms of time of ejection, of $(125.2-30.9)=94.3$
 days, as inferred from Section~\ref{dirmilli}. It is found necessary
 to introduce a constant core flux density, of value comparable to
 that of the plasmons, to simulate the data. The result of this
 exercise is shown in Figure~\ref{plotratio}. Note that corresponding
 to Epoch 2, the second plasmon has just begun to initiate a rise in
 the SW curve, and again at Epoch 5, the same phenomenon is
 observed. Hence the discontinuities in Figure~\ref{linear} are seen
 to be a consequence of the observed appearance of a new plasmon in
 the SW image, as was stated in Section~\ref{microjet}.

\section{A massive binary black hole system for a central engine?}

A popular model for Active Galactic Nuclei (AGNs) with precessing jets
is that they host at their centres massive binary black hole systems
(Begelman, Blandford \& Rees 1980, Roos 1988, Roos, Kaastra \& Hummel
1993, Villata et al. 1998, Villata \& Raiteri 1999, Romero et
al. 2000, Romero, Fan \& Nuza 2003, Rieger 2004, Maness et
al. 2004). The system under study in the present paper lends itself to
such a possibility. The precession period intrinsic to PKS1830-211 has
been derived from Section~\ref{evaltwo} to be about 30.8 years; as was
the case for 1928+738 (with a jet precession period of 2.9 years; Roos
et al. 1993) and 3C 273 (with an observed jet precession period of 16
years; Romero et al. 2000), this is too rapid to be due to geodetic
precession of the jet-emitting black hole (Begelman, Blandford \& Rees
1980), without suggesting a lifetime ($\sim 10^3$ years) for
gravitational collapse of the binary system that is too short.  One
may infer, however, that the precession period is actually a result of
the orbital motion of the black hole binary (as did Roos et
al. 1993). In this case, the jet velocity is modulated by the orbital
motion of the jet-emitting black hole. If $P$ is the precession
period, $G$ the gravitational constant, and $m$ and $M$ the masses of
the two black holes, then, with $r$ as their separation from each
other:
\begin{equation}
r^3 = {P^2 \over {4\pi^2}}G(m+M)
\end{equation}

\noindent For a precession period in units of 30.8 years, and scaling
the masses by $10^8\;{\rm M}_\odot$ and the separation by $10^{16}$
cm,
\begin{equation}
r_{16}=6.833(P_{30.8})^{2/3}(m_8 + M_8)^{1/3}\;\;{\rm cm}
\label{eq:gravrad}
\end {equation}
\noindent The gravitational lifetime of the system is given by:
\begin{equation}
t_{grav}=2.9\times10^5{{M_8\over m_8}\; r_{16}^4 \over(1+ {m_8\over M_8}) M_8^3} \;\;{\rm yrs}
\end{equation}
\noindent In this case, using Equation~\ref{eq:gravrad}, the lifetime of the system 
until gravitational collapse is:
\begin{equation}
t_{grav}=6.322 \times 10^8 {\phd\phd(M_8 + m_8)^{1/3} \over m_8M_8}\;\;{\rm yrs}
\end{equation}

The motion of the core of the jet-emitting black hole, orbiting around
its companion, would be on the scale of microarcseconds, undetectable
in the 43 GHz observations used in this work.

\section{Discussion and Conclusions}

The observations of Jin et al.(2003) permit an analysis of the blazar
source of PKS 1830-211 under atypical circumstances. Firstly, changes
in the core of the source on scales of tens of microarcseconds could
be probed because of the fact that the source is doubly imaged within
a {\it single} VLBI field of view, and hence uncertainties in the
separation of the two core images could be restricted to essentially
thermal noise limits. Secondly, there is some degree of magnification
due to lensing, which, while its numerical value cannot be determined
without the existence of a standard candle/ruler within the source,
still provides a closer look than might otherwise have been possible
(a typical linear magnification factor would be about 3). Thirdly,
this lensed system permits a detailed study of a very distant blazar
(redshift $z_s=2.51$). These observations, coupled with data from
observed changes in the milliarcsecond scale of structure, make it
possible to estimate the precession period of the relativistic jet in
the blazar source in PKS1830-211. The observed period turns out to be
1.08 years, assuming the jet precession axis to be perfectly aligned
with the line of sight to the source.  This is one of the highest
redshifts for which a blazar jet's precession period has been actually
measured.  For PKS1830-211, the typically assumed mass of $\sim 10^8
{\rm M}_{\odot}$ or less for each element of the central powering
black hole system appears to pose no particular evolutionary problems
at this redshift ($cf.$, e.g., Yoo \& Miralda-Escud\'{e} 2004).

\begin{figure}
\begin{center}
\epsfig{file=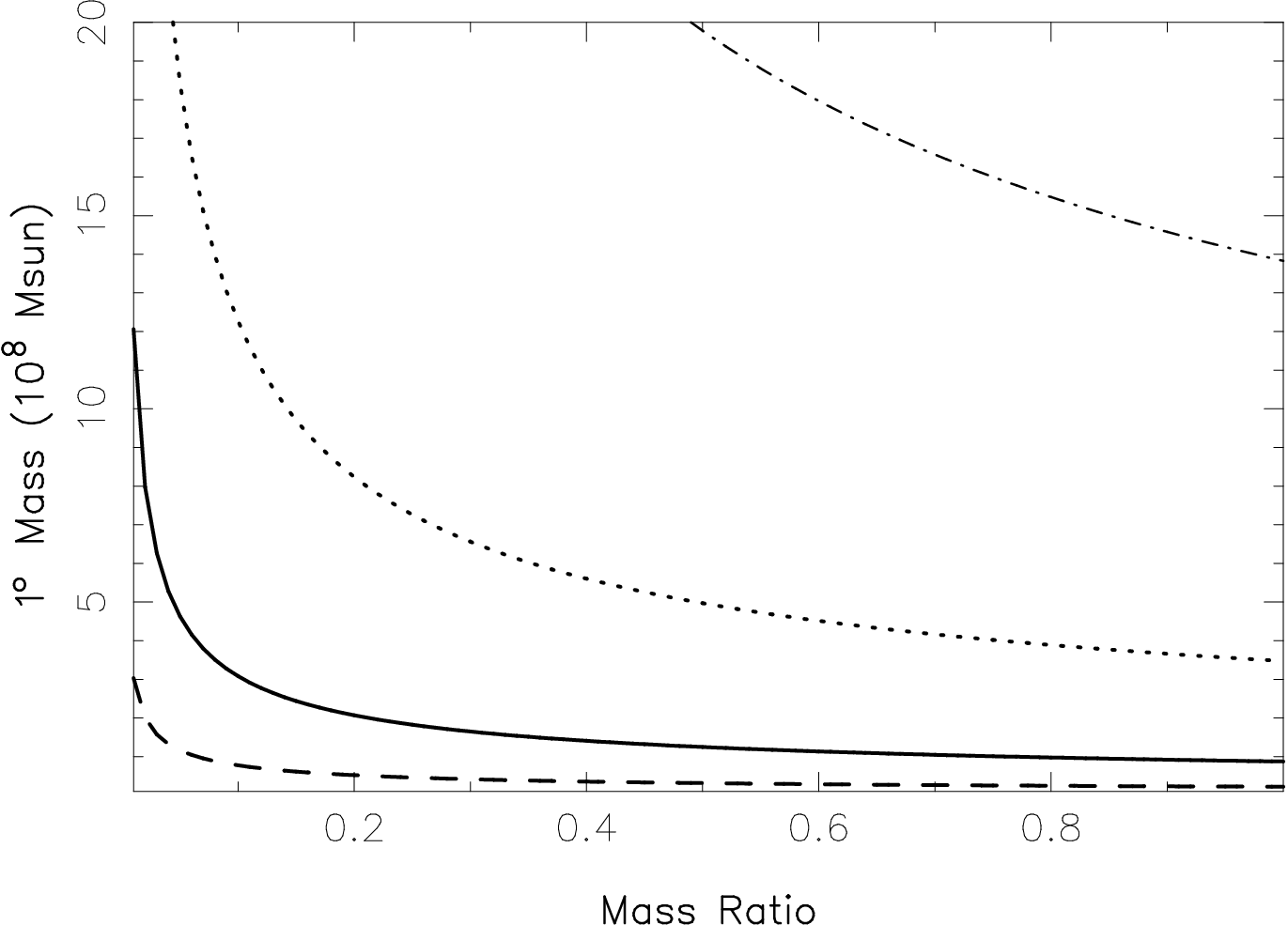,width=6.0 cm,angle=-90}
\end{center}
\caption{\label{binblack}}{Plots of the gravitational lifetime of
PKS1830-211 as a function of primary mass in the inferred binary black
hole system (ordinates, in units of $10^8$ solar masses) and the mass
ratio of the two black holes (abcissae). The plots are for
gravitational lifetimes of {(\it from top down)} $10^7$, $10^8$,
$10^9$ \& $10^{10}$ years respectively.  These plots use the derived
intrinsic period of precession of $30.8$ years; this assumes a jet
bulk velocity of $0.99c$.  }
\end{figure}

 Estimating the mass of a possible central black hole system from a
measured SED is confounded by the fact that PKS1830-211 is seen
through the Galactic plane and is subject to much obscuration in the
optical. Moreover, it is lensed, so there remains the possibility of
spectral changes due to different scales of the source being magnified
to different extents on account of microlensing by stars in the lens
galaxy. However, a method such as measuring the timescale of
variability of gamma ray emission from the source can be
used. Although two timetracks, one time-delayed with respect to the
other by 26 days, would be observed together, the timescale of
variability is expected to be typically of the order of a day or less
(Liang \& Liu 2003), making it unnecessary to correct for the time
delayed track. Then, following the analysis of Liang \& Liu (2003), an
upper limit to the mass of the jet-emitting black hole can be
derived. (Note that these authors do in fact list a black hole mass of
$10^{9.2}\;\;{\rm M}_{\odot}$ and a calculated minimum variability
timescale of $10^{4.5}$ s in Table 1 of their paper, for the system
1830-210, based on an incorrect source redshift of 1 and the observed
EGRET gamma ray flux, which would need to be corrected for lensing;
the absolute magnification factor is an unknown. If it is taken as a
typical value of 10, one obtains a black hole mass of
$10^{8.7}\;\;{\rm M}_{\odot}$ for the source at the observed redshift
of 2.51; note also that this mass is determined under the assumption
that the gamma ray flux is isotropic or unbeamed. This mass
corresponds to a gravitational lifetime of the binary black hole
system of order $10^8$ years ($cf.$ Figure~\ref{binblack})).

In passing, we note that it has been assumed in the present work that
the plasmons follow straight line paths along the surface of the cone
formed by the jet's precessing nozzle (following from ballistic
ejection of plasmons). On the milliarcsecond scale, the essentially
two epochs' worth of data on the source represented by the single
observation of 14 July 1996 (Garrett et al. 1997), which has been used
here, permits no investigation of a possible curvature in plasmon
motion. Multiepoch radio VLBI studies at, say, 43 GHz, during an
active phase for the system, should be able to shed some light on the
question of whether we are really seeing ballistically ejected
plasmons, or whether the helical appearance of the jet is accompanied
by motion of the observed features along helical paths, in which case
by the time the jet has reached milliarcsecond scales, observational
evidence of curvature in the motion of some features might be
expected. More sensitive than the observed motions of individual
features perhaps would be information from the observed light curves
in each image. Features moving along helical paths would show variable
relativistic beaming if the jet axis is even slightly inclined to the
line of sight ($cf.$, e.g., Schramm et al. 1993), exhibiting a
quasi-periodicity in the intensity of emission.  (Note, however, that
the strict linearity apparent in Figure~\ref{linear} spanning three
and four epochs is a more reliable indication that on the scale of
tens of microarcseconds at least, the motion of the plasmons in the
present study show linear velocities, with little sign of curvature).

Lastly, there is a question regarding the plasmon lifetimes. Are the
plasmons we `see' in Jin et al. (2003) on the scale of tens of
microarcseconds likely to be the progenitors of features such as have
been seen on the scale of milliarcseconds in Garrett et al. (1998)? We
think this is unlikely; longer-lived plasmons would not permit a
simple analysis such as has been possible in this work, with almost
single plasmon events to be observed on the scale of tens of
microarcseconds. It would appear that the plasmon emission mechanism
is of a highly variable nature, only occasionally producing long-lived
plasmons such as were seen in Garrett et al. (1998), but for most part
producing relatively short-lived plasmons in a quiescent phase.

\begin{table}
\caption{\label{epochdata}}{Data from the observations of 14 July 1996 (Garrett et al.1998)}
\begin{tabular}{lcc}
\hline
Feature & Flux Density& Position w.r.t. core image \\
&(mJy)&($\Delim_{\alpha,A\; or\; B}, \Delim_{\delta,A\; or\; B})$(mas)\\
\hline
A1 (core) & $191.7 \pm 1.4 $& $(\phm 0.000, \phm 0.000)$ \\
A2        & $58.9 \pm 1.5$ & $(-0.269 \pm 0.002,\phm 0.299 \pm 0.002)$\\
A3        & $8.9 \pm 1.4$  & $(\phm 0.154 \pm 0.013, \phm 0.663 \pm 0.012)$\\
A4 (knot) & $36.4 \pm 1.4$ & $(\phm 0.980 \pm 0.003, -0.150 \pm 0.003)$\\
A5        & $19.7 \pm 1.5$ & $(\phm 1.336 \pm 0.007, -0.334 \pm 0.007)$\\
A6        & $11.6 \pm 2.6$ & $(-0.582 \pm 0.033,-1.502 \pm 0.038)$\\
\hline
B1 (core) & $189.2 \pm 1.6$ & $(\phm 0.000, \phm 0.000)$ \\
B2        & $19.5 \pm 1.5$ & $(\phm 0.279 \pm 0.005, -0.079 \pm 0.006)$\\
B3        & $24.1 \pm 2.4$ & $(-0.266 \pm 0.012, \phm 1.008 \pm 0.016)$\\
B4 (knot) & $18.7 \pm 1.4 $& $(-0.319 \pm 0.005, \phm 0.655 \pm 0.005)$\\
B5        & $43.6 \pm 3.1$ & $(-0.536 \pm 0.015, \phm 0.352 \pm 0.009)$\\
B6        & $16.4 \pm 1.5$ & $(-0.159 \pm 0.007, -0.344 \pm 0.007)$\\
\hline
\end{tabular}
\end{table}

\section{Acknowledgements}

SN wishes to acknowledge Dipankar Bhattacharya and Shiv Sethi for
being reliable sounding boards, especially with regard to numerical
methods, and Vivek Dhawan for raising a couple of critical
questions. We thank an anonymous referee for helpful comments on this
work, which led to improvements.

\section{Appendix: The Mixed Adjustment Model}

From Section~\ref{milli}, the five constraint equations {\bf F},
Equations~\ref{eq:cons1}, \ref{eq:cons2} and the three implicit in
equation~\ref{eq:cons3}, link the four unknowns, the transformation
matrix elements $T_i$ $(i=1...4)$ (constituting vector {\bf X} in the
present analysis), with sixteen observational quantities (positional
separations, from their respective core images, of the eight features
A4(B4), A2(B2), A5(B5) and A6(B6), as seen in both the NE and the SW
images.  These sixteen quantities constitute the vector {\bf L} in
this section.  The positions of the observational features, with
errors of fit, are listed in Table~\ref{epochdata}.  The five
equations {\bf F} overdetermine the quantities in {\bf X}, hence a
numerical method, the Mixed Adjustment Model, is employed to solve for
{\bf X}. Note that the observations and the unknowns are implicitly
related. The treatment here follows that described in A.Leick (1995),
Section 4.4.

Let in the course of the application of this method, the 16 adjusted
observations be denoted by {\bf L$_a$}, and the adjusted vector of
unknowns be denoted by {\bf X$_a$}.  The mathematical model is then:
\begin{equation}
{\bf F({\rm \bf L}_a, {\rm \bf X}_a)} = 0
\label{eq:mathmod}
\end{equation}

\noindent Let {\bf X$_0$} be a set of known but approximate values of
{\bf X}, and denote the vector of observations by {\bf L$_b$}. Then
Equation~\ref{eq:mathmod} may be written as:
\begin{equation}
{\bf F({\rm \bf L}_b+{\rm \bf V},{\rm \bf X}_0+{\rm \bf X})}= {\bf 0},
\end{equation}
\noindent where {\bf V} and {\bf X} are the residuals at each stage of
iteration. This essentially nonlinear form is linearized about the
point ({\bf L$_b$, X$_0$}):
\begin{equation}
\label{eq:linmod}
\php_5{{\rm \bf B}_{16}}\php_{16}{\rm \bf V}_1+ \php_5{\rm \bf A}_4\php_4{\rm \bf X}_1 + \php_5{\rm \bf W}_1={\bf 0},
\end{equation}

\noindent where:
\begin{equation}
{\php_5{\rm \bf B}_{16}}=  {\partial{\rm \bf F}\over {\partial{\rm \bf L}}}\mid_{ \php_{({\rm \bf X}_0,{\rm \bf L}_b)}}
\end{equation}

\begin{equation}
{\php_5{\rm \bf A}_{4}}=  {\partial{\rm \bf F}\over {\partial{\rm \bf X}}}\mid_{ \php_{({\rm \bf X}_0,{\rm \bf L}_b)}}
\end{equation}

\begin{equation}
{ \php_5{\rm \bf W}_1}=  {\rm \bf F}({\rm \bf L_b},{\rm \bf X_0})
\end{equation}

\noindent Now a least-squares estimate of {\bf X} is sought, with the
target function {\bf V}$^T${\bf PV}, where {\bf P} is the $16\times
16$ weight matrix constructed from the covariance matrix of the
observations, $\sigma_o^2 \Sigma^{-1}_{\bf \rm L_b}$, where
$\sigma_o^2$ is the {\it a priori} variance of unit weight.  The
target function is minimized subject to the constraints imposed by the
linearized mathematical model, Equation~\ref{eq:linmod}, with a vector
of Lagrangian multipliers, {\rm \bf K}. Hence, one minimizes with
respect to {\bf V}, {\bf K} and {\bf X}, the function:
\begin{equation}
\label{eq:targetfn}
\phi({\rm \bf V},{\rm \bf K},{\rm \bf X})={\rm \bf V}^T{\rm \bf PV} - 
2{\rm \bf K}^T({\rm \bf BV}+{\rm \bf AX}+{\rm \bf W}),
\end{equation}
\noindent and the three constraint equations arising from this process
are used to calculate {\bf V},{\bf K} and {\bf X}.  Since the actual
mathematical model is non-linear, it is necessary to iterate to a
solution the actual values of {\bf V}, {\bf K} and {\bf X}. The
iterative process is said to converge if:
\begin{equation}
\mid({\rm \bf V}^T{\rm \bf PV})_i -({\rm \bf V}^T{\rm \bf PV})_{(i-1)}\mid < \epsilon,
\end{equation}
where $\epsilon$ is a small positive number (here taken as
$10^{-8}$). In the present case, the calculation converged in five
iterations, with a final value of {\bf V}$^T${\bf PV}=9. The estimated
{\it a posteriori} variance of unit weight, ${\sigma}_o^2$, is
therefore {\bf V}$^T${\bf PV}/(5-4)=9.

\end{document}